\documentclass[12pt,a4paper]{jpconf}
\usepackage[latin1]{inputenc}
\usepackage{amsmath}
\usepackage{amsfonts}
\usepackage{amssymb}
\usepackage{graphicx}
\usepackage{iopams}
\usepackage{textcomp}
\usepackage{siunitx}
\usepackage{lineno}
\newcommand{\ptt}{\ensuremath{p_{\rm T}}}
\newcommand{\wpm}{\ensuremath{\rm W^{\pm}}}

\begin{document}
\title{Vector-boson production in p--Pb collisions with ALICE at the LHC}
\author{Kgotlasele Johnson Senosi, for the ALICE collaboration}
\address{Department of Physics, University of Cape Town, Private Bag X3, 7701 Rondebosch}
\address{Department of Nuclear Physics, iThemba Laoratory of Accelerator Based Sciences, Old Faure Road, Faure, Somerset West, South Africa}

\begin{abstract}
ALICE (A Large Ion Collider Experiment) is designed and optimized to study ultra-relativistic heavy-ion collisions at the LHC, in which a hot and dense, strongly-interacting medium is created. Vector bosons (W and Z) are produced in hard scattering processes and interact weakly with the medium formed in heavy-ion collisions. Thus, they present a suitable reference for processes which are heavily affected by the medium. In proton-nucleus collisions their production can be used to study the modification of parton distribution functions in the nucleus and to test the validity of binary-collision scaling for hard processes. The production of W and Z bosons is studied in p--Pb collisions at a centre-of-mass energy of $\sqrt {s_{\rm NN}}$ = 5.02 TeV with the ALICE muon spectrometer at forward ($2.03 < \mathit{y}^{\mu}_{\textrm{cms}} < 3.53$) and backward ($-4.46< \mathit{y}^{\mu}_{\textrm{cms}} <-2.96$) rapidity. W bosons are studied via the inclusive single muon differential $p_{\rm T}$ spectrum whereas the Z-boson signal is observed in the invariant mass distribution of unlike-sign muon pairs as a peak around the Z-boson mass. In this presentation the measured cross sections of W and Z bosons and the W-boson yield per centrality interval will be discussed. The cross-sections are compared to theoretical calculations.
\end{abstract}

\section{Introduction}
The high energies available in hadronic collisions at the Large Hadron Collider determine an abundant production of hard probes, such as heavy quarks, quarkonia, high-\ptt\ jets and intermediate vector bosons (\wpm\ and $\mathrm{Z^{0}}$). The large masses of W (80 GeV/$c^{2}$) and Z (91 GeV/$c^{2}$) bosons imply that they are formed during initial hard parton scattering processes with a formation time of about 0.003 fm/$c$ and the mean decay time, which is inversely proportional to their width, is about 0.09 fm/$c$ \cite{PhysRevD.86.010001}. Precise theoretical predictions of the W- and Z-boson cross sections in proton-proton collisions make them good standard candles for luminosity measurements. In addition, they can be used to constrain parton distribution functions (PDFs) at high momentum transfer (\textit{Q}) equal or larger than the mass of the vector bosons. In high-energy heavy-ion collisions where nuclear matter undergoes a phase transition to a deconfined state called Quark-Gluon Plasma (QGP), these electroweak bosons decay either before or during the formation of the QGP. Vector bosons and their leptonic decay products do not interact strongly and thus they should not be affected by the QGP. Therefore, they give access to the initial-state properties in nuclear collisions and their yields provide a benchmark for the binary collision scaling of hard processes.
Finally, the isospin dependent production of W and Z bosons and their weak coupling nature makes them good probes to study initial-state effects (for example, the nuclear modification of PDFs and isospin effects) in proton-lead and lead-lead collisions \cite{Paukkunen:2010qg}. 

\section{Experimental apparatus and data sample}
\label{exp}
The ALICE experiment~\cite{alice, Abelev:2014ffa} is an ensemble of various detectors each with a specific purpose. The V0 scintillators, the second layer of the silicon pixel detector (SPD) and the zero degree calorimeters (ZDC) are used to characterize events according to their activity (measure of centrality). Event activity estimators V0A(C) and CL1 associated with V0-A(C) and SPD classify events according to the charged-particle multiplicity, whereas ZNA(C) uses the neutron energy deposited in the ZDC. The events are divided into four equal bins of event activity corresponding to 20\% fraction of events  (0-20\%, 20-40 \%, 40-60\% and 60-80\%). The V0 scintillators, covering the pseudorapidity ranges $2.8 < \eta_{\rm lab}\ < 5.1$ (V0A) and $-3.7 < \eta_{\rm lab}\ < -1.7$ (V0C) are also used for triggering. The SPD second layer covers the $|\eta_{\rm lab}| < 0.9$ pseudorapidty region. The zero degree calorimeters (ZNC and ZNA) are located 112 meters away on either side of the interaction point along the beam pipe.      
\linebreak
The muon spectrometer has a forward pseudorapidity coverage of $-4.0 < \eta_{\rm lab} < -2.5$ and consists of an absorber, five tracking stations, a dipole magnet, a muon filter and two triggering stations. The absorber reduces the background from pions and kaons, such that mostly muons enter the tracking system where their transverse momentum (\ptt) is determined. Furthermore, the muon filter, placed between the tracking and trigger stations, stops remaining background hadrons and low momentum muons ($\ptt<0.5$ GeV/\textit{c}) to ensure high purity of the triggered muon-candidate sample. In addition, the triggering stations provide an approximate \ptt\ measurement which is employed for the trigger decision.\\
\linebreak
The results described here are based on a data sample collected in proton-lead (p--Pb) collisions at $\sqrt{s_{\rm NN}} = 5.02$ TeV ($E_{\rm proton} = 4$ TeV and $E_{\rm lead} = 1.58$ /$A$ TeV) in which two beam configurations were used, protons going towards the spectrometer (p--Pb) and vice-versa (Pb--p). The asymmetry in energy translates into the centre-of-mass system boost of $\Delta y = 0.465$ rapidity units in the proton direction. Thus, the rapidity intervals covered by the muon spectrometer are, $2.03 < y_{cms}^{\mu} < 3.53$ (p--Pb) and $-4.46 < y_{cms}^{\mu} < -2.96$ (Pb--p), respectively, where the direction of the proton defines positive rapidities.  Henceforth, these will be respectively referred to as forward and backward rapidity.  
The data samples consist of events collected with minimum bias (MB) trigger in coincidence with high-\ptt\ single-muon (MSH) and low-\ptt\ di-muon (MUL) trigger, for W and Z bosons respectively. The MB trigger is defined as the coincidence of signals in both V0 detectors. The single muon trigger is a coincidence between a MB trigger and the presence of a muon track in the triggering station with $\ptt \gtrsim $ 4 GeV/\textit{c}. The di-muon trigger is a coincidence of MB trigger and an un-like sign muon pair of $\ptt \gtrsim $ 0.5 GeV/\textit{c}. The single and di-muon triggered data samples are used for W and Z bosons analysis respectively. The integrated luminosities were measured to be 5.01$\pm$0.20 $\rm nb^{-1}$ (p--Pb) and 5.81$\pm$0.2 $\rm nb^{-1}$ (Pb--p) \cite{Adam:2015jsa}. \\


\section{Analysis strategy}
The muon tracks are required to be in the detector acceptance $-4.0 < \eta_{\rm lab}\ < -2.5$ and to come from a reconstructed interaction vertex. Another geometrical cut taken into consideration is the angle of the tracks at the end of the absorber covering the range $170\si{\degree} < \theta_{\rm abs} < 178\si{\degree}$, which rejects particles crossing non-uniform material sections at the absorber edges. In order to remove the background from interactions between beam particles and the residual gas, an additional cut \textit{p}$\times$DCA based on the product of the muon track momentum and its transverse distance of closest approach to the interaction point, is used to remove tracks which do not point to the interaction vertex.
\subsection{W boson}
\begin{figure}[!h]
\centering
 \includegraphics[width=0.3\textwidth]{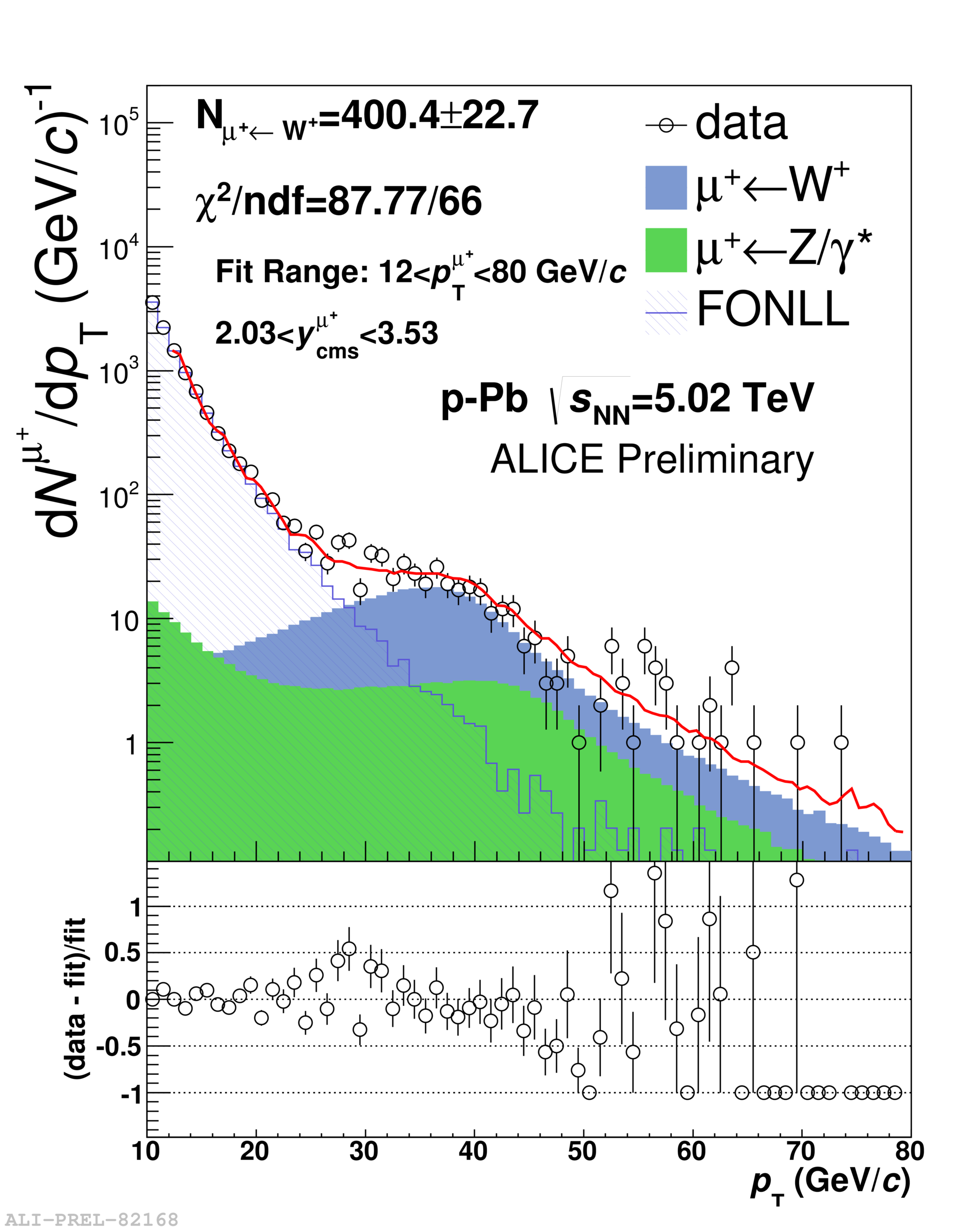}
 \includegraphics[width=0.3\textwidth]{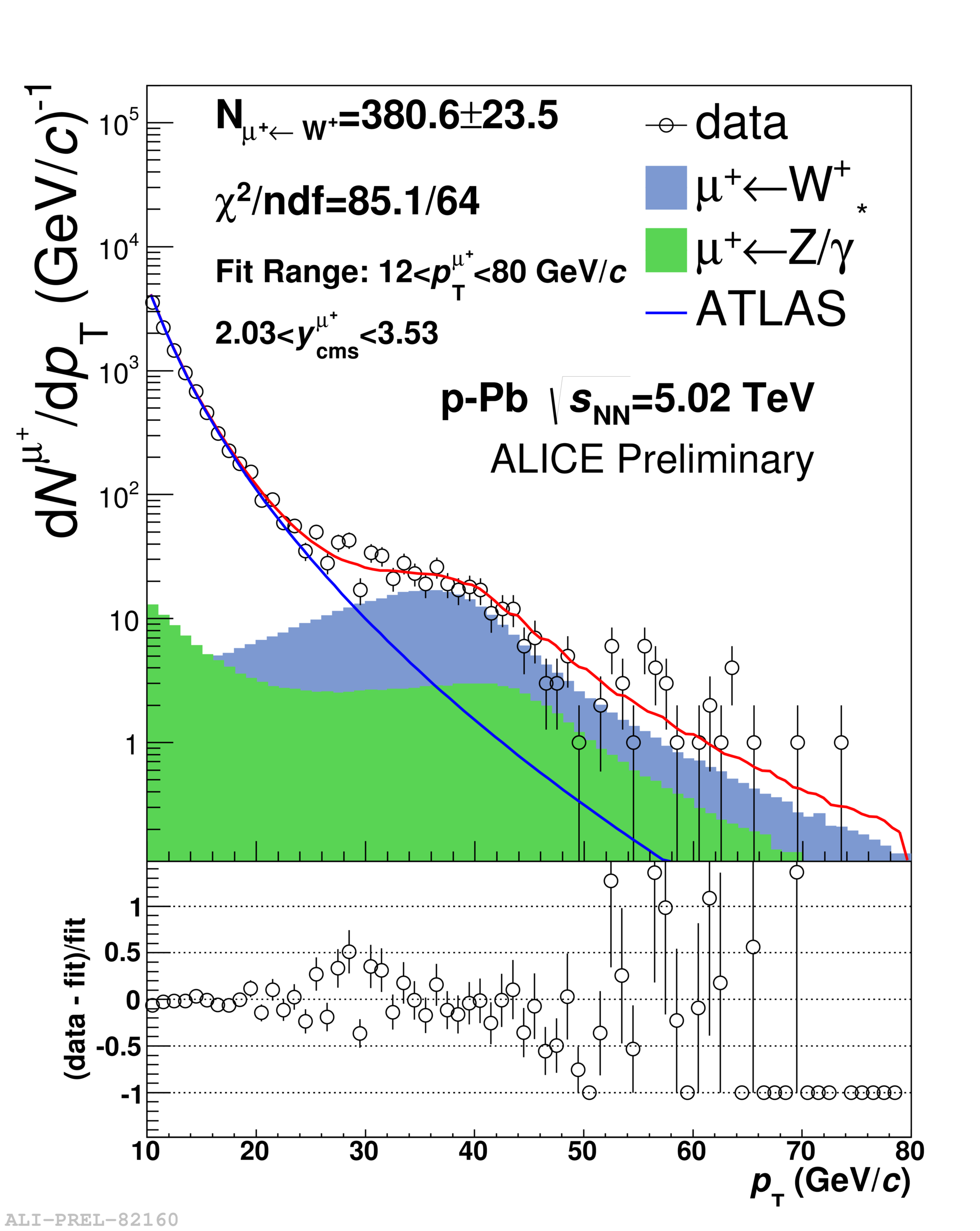}
 \includegraphics[width=0.3\textwidth]{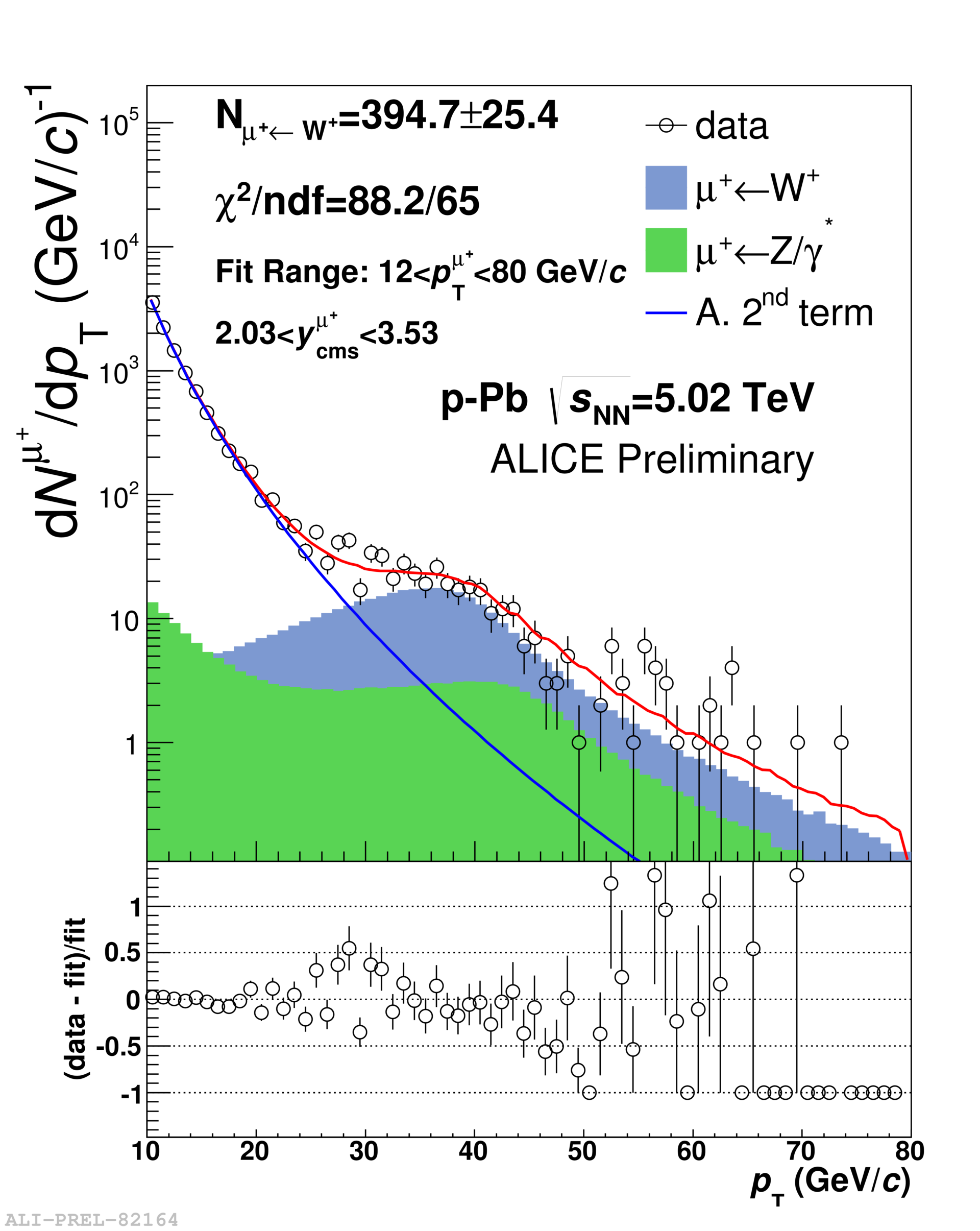}
 \caption{Examples of the combined fit to the raw single muon \ptt-spectrum at forward rapidity in the case of positive muons. The errors on the data are statistical. Heavy-flavour templates based on FONLL \cite{fonll} (left), ATLAS function \cite{atlas} (middle) and the 2$^{\rm nd}$ term (second addend) of the ATLAS function (right).}
 \label{fits1}
\end{figure}

The analysis is based on the extraction of the W-boson decay contribution to the total single muon \ptt\ spectrum. The semi-muonic decays of W bosons form a Jacobean peak with a maximum around $\ptt\ \sim M_{\rm W}/2$. Muons from Z-boson ($\rm Z^{0}/\gamma^{*}$) decays are dominant source of background above $\ptt\ \sim 35$ GeV/\textit{c} whereas muons from decays of heavy-flavour hadrons (HF) dominate the lower \ptt\ region ($10 < \ptt\ < 35$ GeV/\textit{c}) \cite{zaida}. The number of W bosons ($N_{\rm W\leftarrow \mu}$) is extracted from the single-muon \ptt\ spectrum using a combined fit composed of suitable templates or functions for the various contributions. The templates for muons from  W and Z bosons are based on the next-to-leading-order (NLO) event generator POWHEG \cite{powheg} and also PYTHIA6.4 \cite{pythia} to take into account the nuclear modification of the PDFs (nPDF). The templates are generated for pp and pn collisions separately to account for the isospin dependence of W- and Z-boson production and are determined for each $y_{\rm cms}$ interval. These templates are then combined into a fit function 
 
\begin{equation}
\label{comb_fit}
 f(\ptt) = N_{\mu \leftarrow \rm HF} \cdot f_{\mu \leftarrow \rm HF} + N_{\mu \leftarrow \rm W}\cdot f_{\mu \leftarrow \rm W} + N_{\mu \leftarrow {\rm Z^{0}}/\gamma^{*}}\cdot f_{\mu \leftarrow {\rm Z^{0}}/\gamma^{*}}
\end{equation}
where $f_{\mu \leftarrow \rm HF}(\ptt)$ is either a phenomenological function \cite{atlas} or the FONLL-based heavy-flavour decay-muon template \cite{fonll} and $f_{\mu \leftarrow \rm W}$ and $f_{\mu \leftarrow {\rm Z^{0}}/\gamma^{*}}$ are W- and $\rm Z^{0}/\gamma^{*}$-boson decay templates. In Equation \ref{comb_fit}, $N_{\mu \leftarrow \rm HF}$ and $N_{\mu \leftarrow \rm W}$ are free parameters whereas $N_{\mu \leftarrow {\rm Z^{0}}/\gamma^{*}}$ is constrained by the ratios of the measured cross sections computed with POWHEG and PYTHIA6.4. The modification of nuclear PDFs, which is not included in POWHEG templates, is taken into account by the EPS09 \cite{eps09} parametrization included in PYTHIA6.4 templates. The number of muons from W-boson decays is extracted between 10 $< \ptt < $ 80 GeV/\textit{c} as a weighted average over fitting trials (varying fit range, background functions, fraction of Z to W bosons and detector alignment). Figure~\ref{fits1} shows examples of the fits to raw single muon spectrum with only statistical errors varying the heavy-flavour background. The resulting uncertainty on the signal extraction varies between 6\% and 24\% depending on the bin of event activity.

\subsection{Z boson}
\begin{figure}[!h]
\centering
 \includegraphics[width=0.4\textwidth]{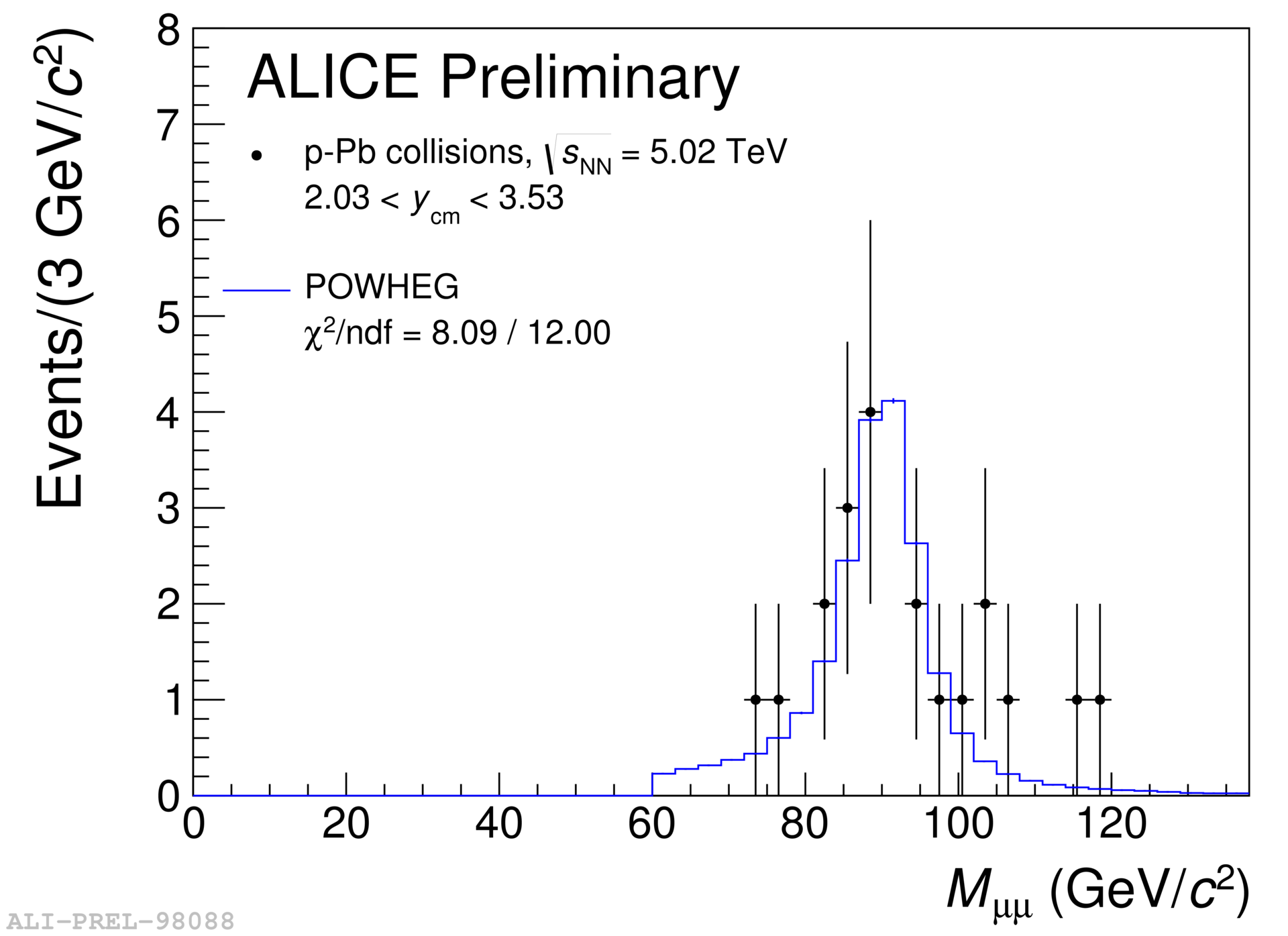}
 \includegraphics[width=0.4\textwidth]{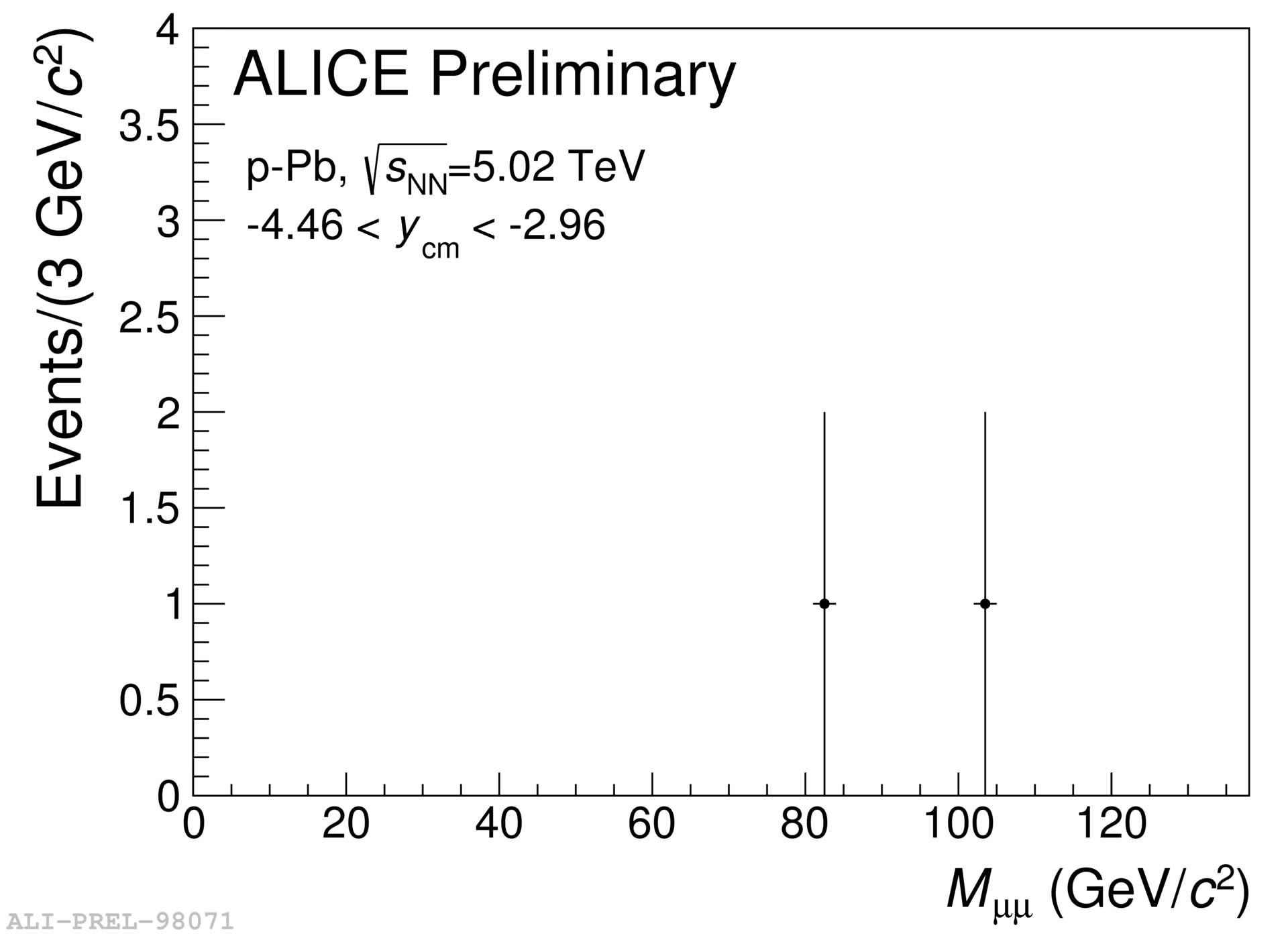}
  \caption{Invariant mass distribution of un-like sign muon pairs with \ptt\ $>$ 20 GeV/\textit{c} in the p-going (left panel) and Pb-going (right panel). The blue line represents the distribution obtained using POWHEG \cite{powheg} simulation and normalised to the number of Z candidates in the data.}
 \label{z_inv}
\end{figure}

The Z-boson signal is a peak on the invariant mass distribution of un-like sign muon pairs. This peak is visible around the mass (90 GeV/$c^{2}$) of the Z boson. The muons are required to pass the cuts detailed in section~\ref{exp}. In addition, each muon is required to have $\ptt>$ 20 GeV/\textit{c} in order to reduce background from lower mass resonances, such as J/$\psi$, $\psi(2S)$, $\Upsilon(1S)$, $\Upsilon(2S)$ and $\Upsilon(3S)$. The resulting invariant mass distribution is shown in Figure~\ref{z_inv}. From the data sample only about 22 (2) candidates with $m_{\mu\mu} > 60$ GeV/$c^{2}$ pass the cuts for p--Pb (Pb--p). The low detector efficiency as well as low acceptance during the Pb--p period led to low statistics. The measured invariant mass distribution (p--Pb) is compared with the expectations for Z-boson detection, tested
by comparing the mean value and width of the distribution with the analogous parameters from Monte Carlo (MC) simulations. The parameters were extracted from a fit using the Crystal-Ball function \cite{Gaiser:1982yw}, which consists of a Gaussian core with power-law tails on both sides. The four tails parameters are fixed to those obtained by fitting MC distribution obtained using POWHEG (NLO) with CT10 and EPS09 (NLO) nuclear PDFs, while the parameters of the Gaussian core are left free. A good agreement of the extracted free parameters is found between data and MC. The contribution of like-sign muons in the region of interest, $60 < m_{\mu\mu} < 120$ GeV/$c^{2}$, was found to be as low as 0.1\%.  

  
Other physics processes, like $Z\rightarrow\tau\tau\rightarrow\mu\mu$, di-muon decay of $t\bar{t}$ and semi-leptonic decays of $c\bar{c}$, $b\bar{b}$ are estimated to contribute 0.7\% and 0.4\% for p--Pb and Pb--p, respectively.


\section{Results}
W- and Z-boson cross sections are computed by correcting $N_{\mu \leftarrow \rm W}$ and $N_{\mu\mu\leftarrow \rm Z}$ with an associated acceptance times efficiency factor (within the signal extraction region) of 88\% (W boson) and 78\% (Z boson) for p--Pb and 76\% (W boson) and 68.42\% (Z boson) for Pb--p, and then normalizing to the integrated luminosity. In Figure 3, the measured cross sections of W bosons ($\sigma_{\mu^{\pm} \leftarrow \mathrm{W}^{\pm}}$) at forward and backward rapidity are compared with pQCD predictions with unmodified CT10 PDFs \cite{ct10} and with modified CT10 PDFs with the EPS09 parametrization \cite{eps09} of nuclear shadowing. Theoretical cross sections with shadowing are in agreement with the measured cross sections within uncertainties.   
\begin{figure}[!h]
\centering
 \includegraphics[width=0.48\textwidth]{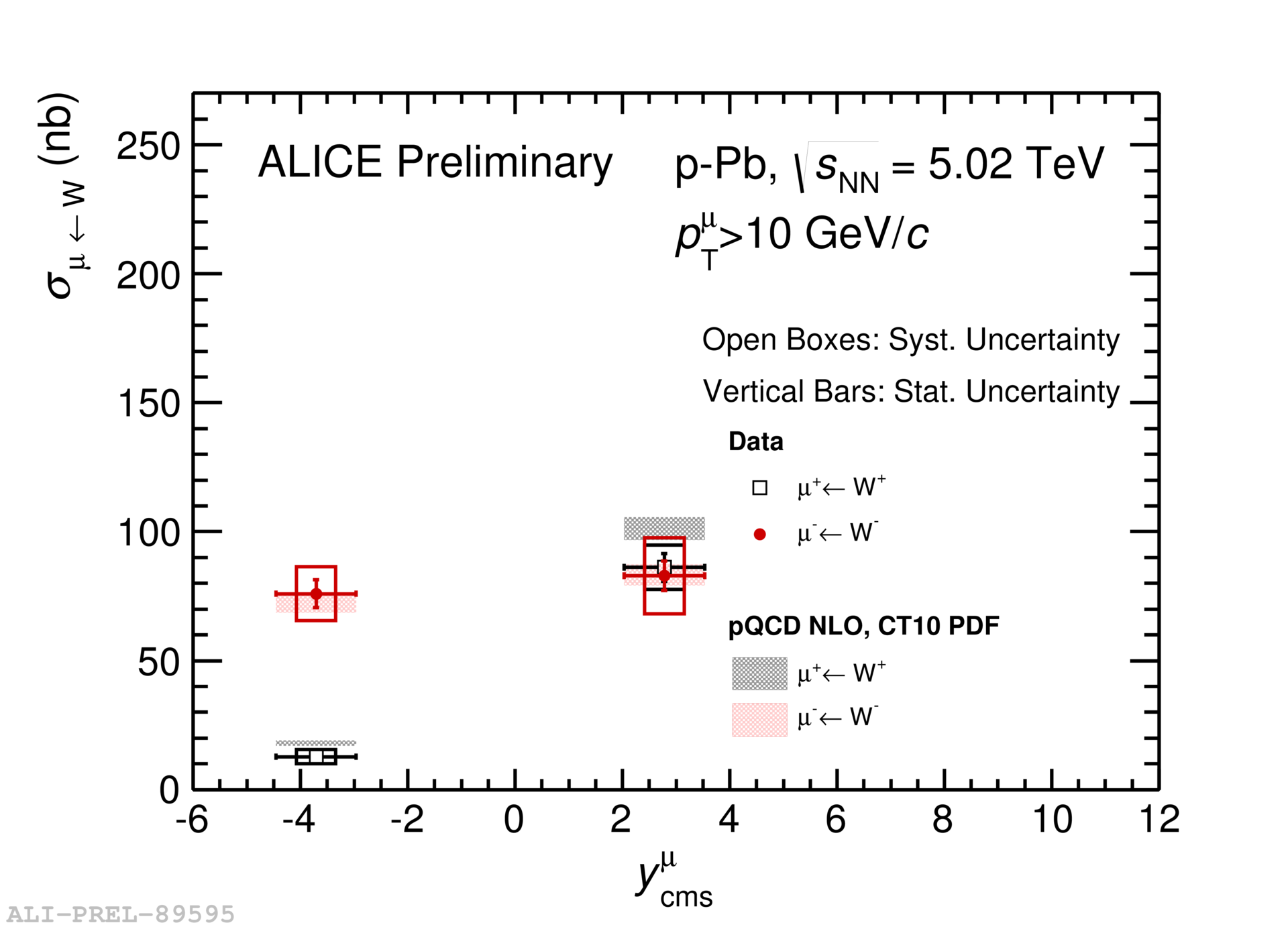}
 \includegraphics[width=0.48\textwidth]{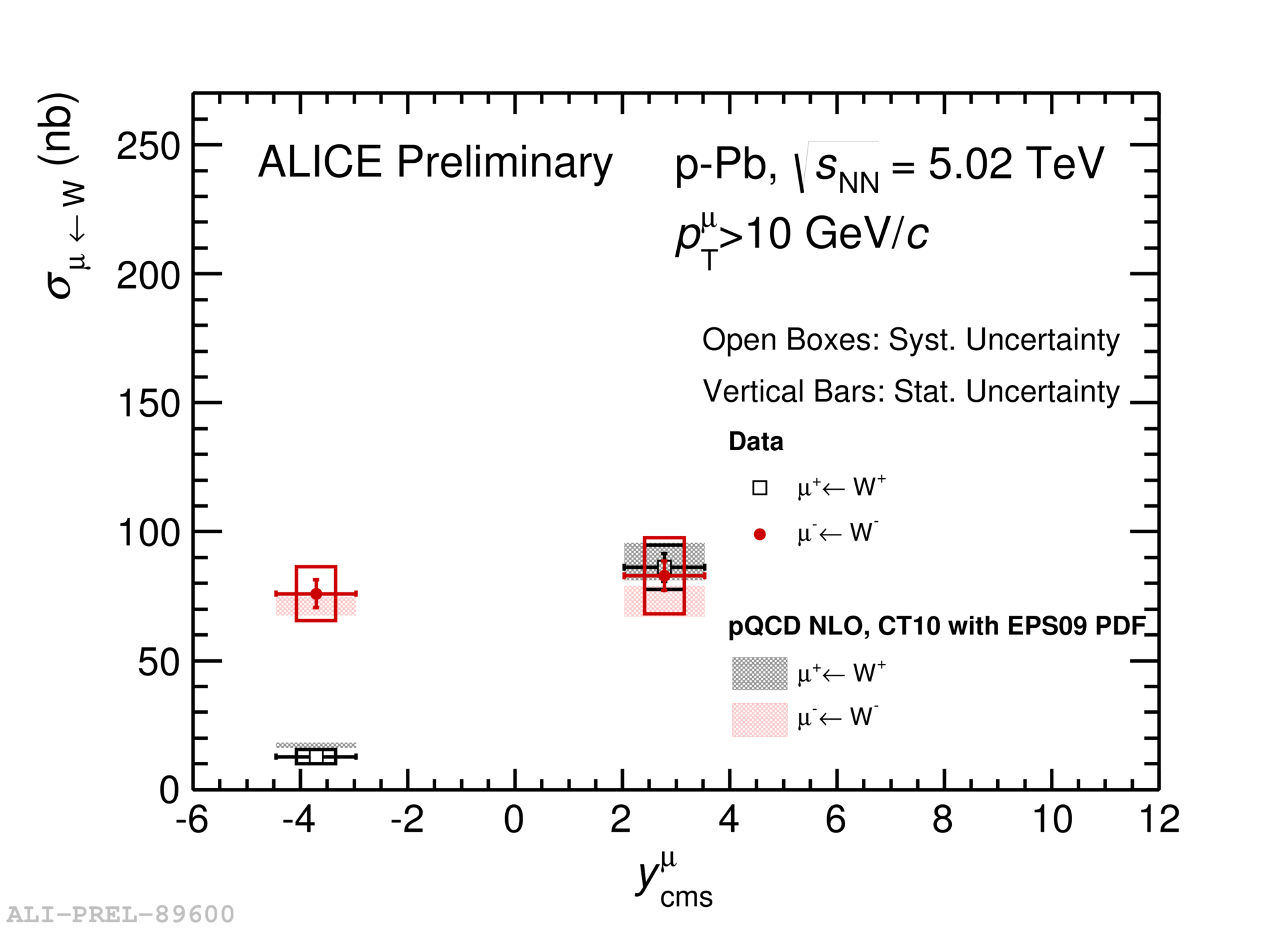}
 \caption{The measured cross section of muons from W-boson compared with pQCD \cite{Paukkunen:2010qg} calculations with CT10 \cite{ct10} PDFs (left) and CT10 PDFs including EPS09 \cite{eps09} nuclear shadowing (right) at forward and backward rapidity. The muons with \ptt$>10$ GeV/$c$ are considered.}
 \label{cross_section}
\end{figure}
In Figure 4 (left), the Z-boson cross sections at forward and backward rapidity are compared with next-to-next-to leading order (NNLO) Fully Exclusive W and Z (FEWZ) \cite{Gavin:2010az} predictions with CT10nlo \cite{Lai:2010vv}, CTEQ6m\cite{PhysRevD.78.013004}, JR09NNLO \cite{JimenezDelgado:2008hf} and MSTW2008NNLO \cite{Martin:2009iq} with and without nuclear PDFs. An upper limit is evaluated for the backward rapidity. The ratios of the measured cross sections (as obtained by ALICE and the LHCb \cite{lhcb}) to FEWZ Z-boson cross sections are shown in Figure~\ref{Z_cross_section} (right). In Figure~\ref{Z_cross_section} (right), the ALICE measurements are in agreement with theoretical calculations in both rapidity intervals.  
\begin{figure}[!h]
\centering
 \includegraphics[width=0.45\textwidth]{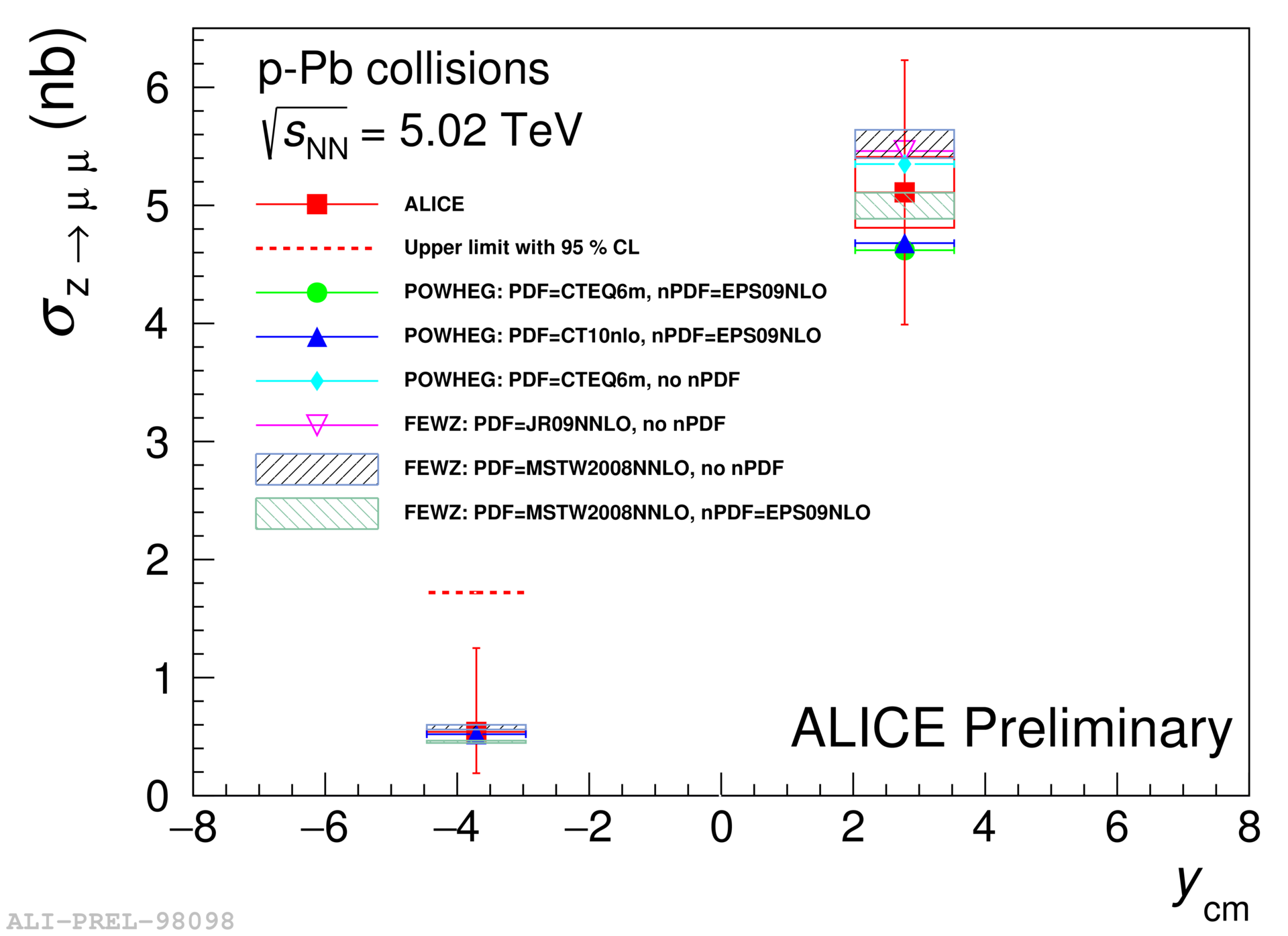}
 \includegraphics[width=0.45\textwidth]{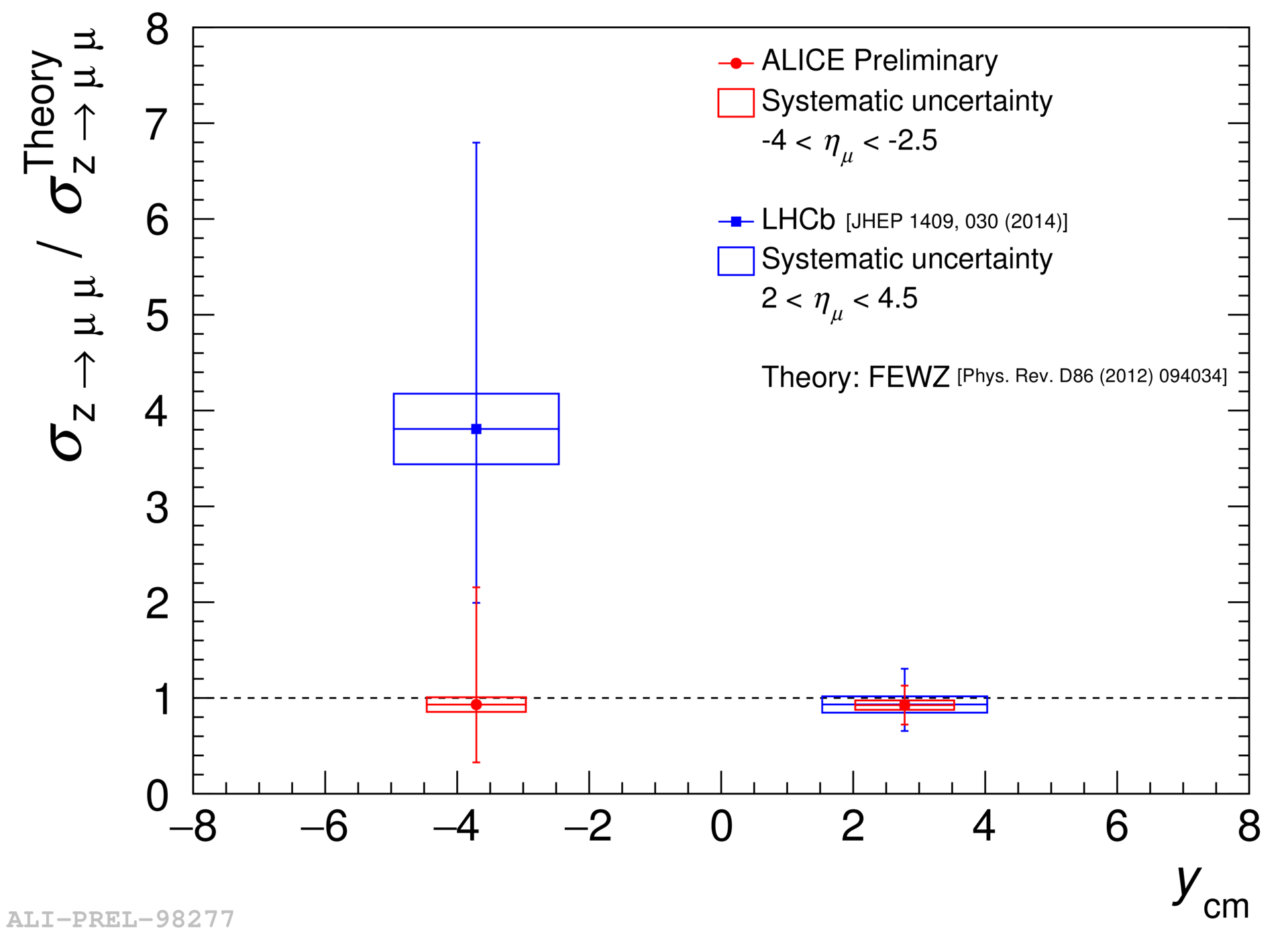}
 \caption{(Left) The measured Z-boson cross section compared to FEWZ \cite{Gavin:2010az} calculations with different PDFs with and without EPS09 \cite{eps09} parametrization of shadowing. (Right) The ratio of the measured Z-boson cross section to that of FEWZ obtained by the ALICE collaboration is compared with those obtained by LHCb collaboration \cite{lhcb}.}
 \label{Z_cross_section}
\end{figure}
\linebreak
W boson and its leptonic decay products do not interact strongly and thus its production is expected to scale with the number of binary  collisions.  This scaling was verified by the CMS \cite{Chatrchyan2012nt,Chatrchyan2014csa} and ATLAS \cite{atlasZ,Aad:2014bha} collaborations for photons, W and Z bosons. In ALICE the binary scaling of W-boson yields is tested by dividing the measured sum of muons from W-boson decays per event by the average number of binary nucleon-nucleon collisions $\langle N_{\rm coll}\rangle$ \cite{centrality}. This measurement is done for minimum-bias events (no event activity selection) and in classes of event activity. The values of $\langle N_{\rm coll}\rangle$ are obtained using methods described in \cite{centrality}. Figure \ref{binary_scaling} shows the yield normalized to $\langle N_{\rm coll}\rangle$ for both backward and forward rapidity. The uncertainties on the ratio Yield$_{\mu \leftarrow \mathrm{W}}$/$\langle N_{\mathrm{coll}}\rangle$ include the uncertainties on $\langle N_{\mathrm{coll}}\rangle$ which varies between 8\% and 21\% depending on the event-activity class. The $\langle N_{\mathrm{coll}}\rangle$-normalized yield is independent of event activity and compatible among estimators within uncertainties.
 
\begin{figure}[!h]
\centering
 \includegraphics[width=0.45\textwidth]{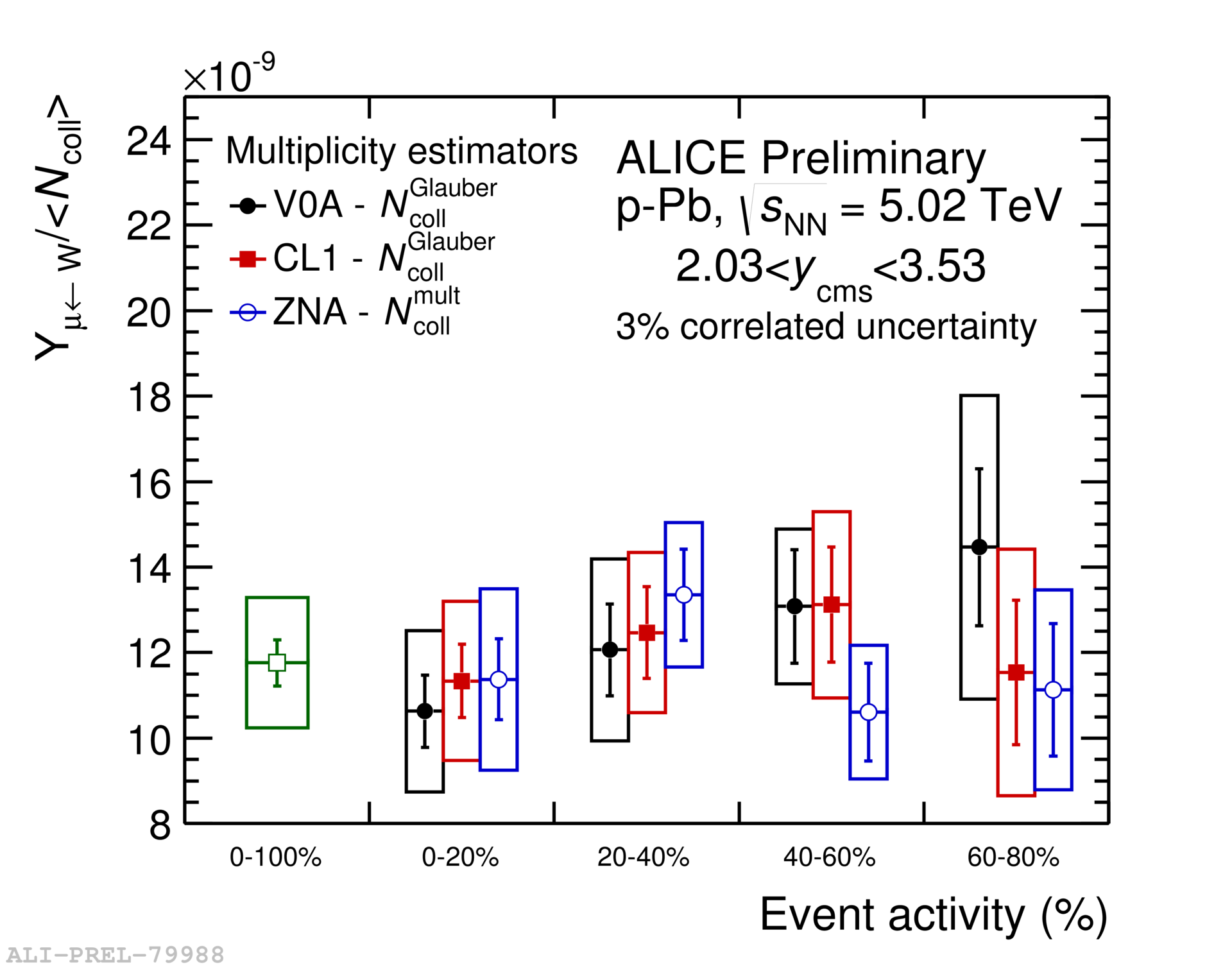}
 \includegraphics[width=0.45\textwidth]{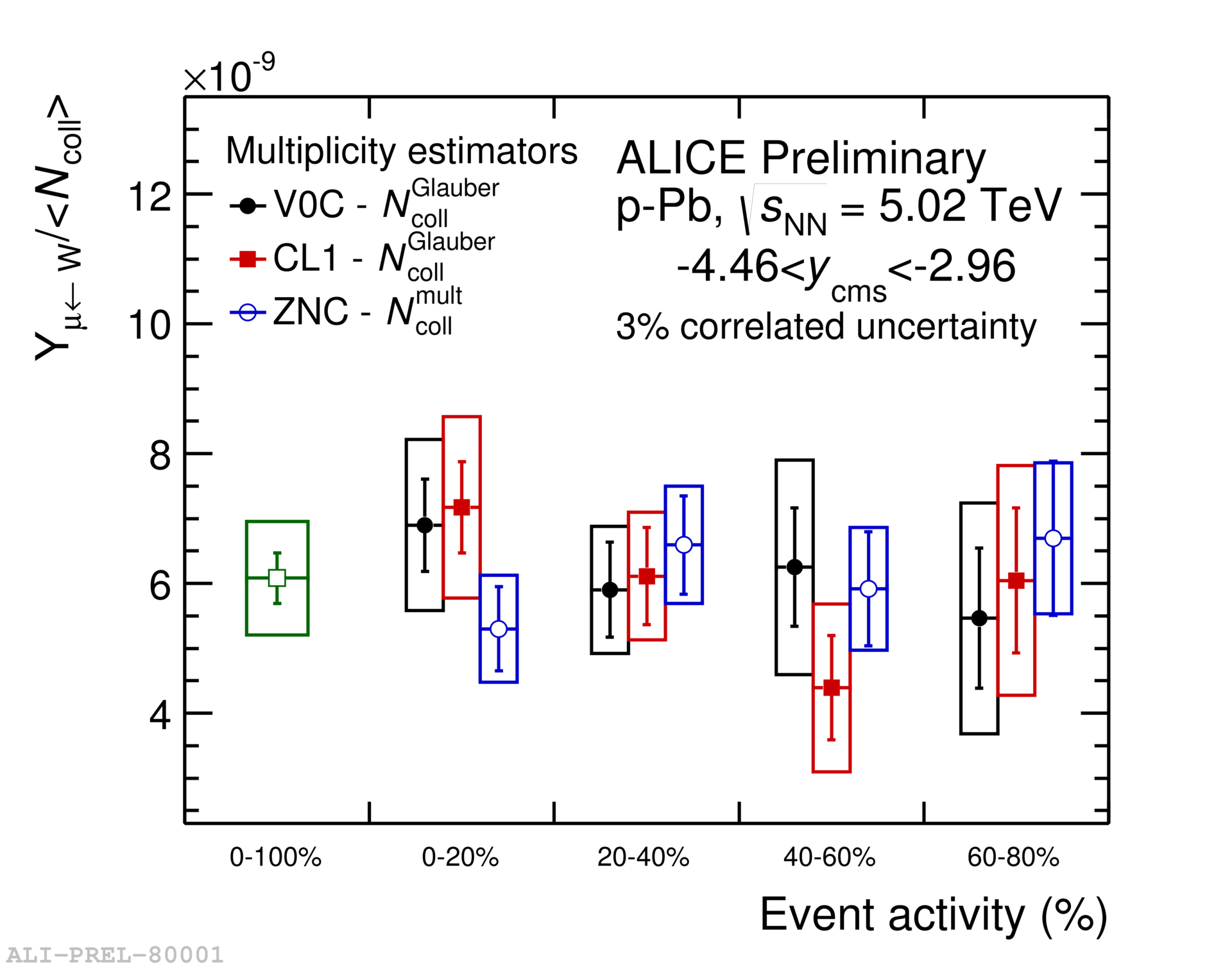}
 \caption{Yield of muons from W decays at forward (left) and backward (right) rapidity  as a function of event acivity. The green points represent the values in minimum-bias collisions.}
 \label{binary_scaling}
\end{figure}

\section{Conclusions}
The production cross sections of W and Z bosons in p--Pb collisions at $\sqrt{s_{\mathrm{NN}}} =$ 5.02 TeV have been measured in two rapidity intervals. The comparison of these cross sections with theoretical predictions based on pQCD \cite{Paukkunen:2010qg} and FEWZ \cite{Gavin:2010az} calculations with normal proton PDFs shows agreement within uncertainties for W and Z, respectively. Taking into account the EPS09 \cite{eps09} parametrization of nuclear shadowing of the PDFs further improves the agreement between the calculations and the data at forward rapidity where shadowing is expected to be important. The measured yield of muons from W-boson decays normalized to $\langle N_{\mathrm{coll}}\rangle$ as function of event-activity shows that the W-boson production scales with $\langle N_{\mathrm{coll}}\rangle$.

\section*{References}

\providecommand{\newblock}{}

\end{document}